\documentclass[letter]{aa} 

\usepackage{graphicx}
\usepackage{txfonts}
\usepackage{hyperref}
\usepackage[para, online, flushleft]{threeparttable}

\newcommand{\Hunit}{km s$^{-1}{\rm Mpc}^{-1}$}

\begin{document} 


   \title{TDCOSMO V: Strategies for precise and accurate measurements of the Hubble constant with strong lensing}
   \titlerunning{Strategies to measure $H_0$ from time delays}

   \author{
Simon~Birrer\inst{1}\fnmsep\thanks{E-mail: sibirrer@stanford.edu}
\and
Tommaso~Treu\inst{2}
}

   \institute{Kavli Institute for Particle Astrophysics and Cosmology and Department of Physics, Stanford University, Stanford, CA 94305, USA
         \and
             Physics and Astronomy Department, University of California, Los Angeles CA 90095, USA
             }

   \date{Accepted XXX. Received YYY; in original form ZZZ}

 
  \abstract
   {Strong lensing time delays can measure the Hubble constant H$_0$ independently of any other probe.
   Assuming commonly used forms for the radial mass density profile of the lenses, a 2\% precision has been achieved with seven Time-Delay Cosmography (TDCOSMO) lenses, in tension with the H$_0$ from the cosmic microwave background. However, without assumptions on the radial mass density profile ---and relying exclusively on stellar kinematics to break the mass-sheet degeneracy--- the precision drops to 8\% with the current data obtained using the seven TDCOSMO lenses, which is insufficient to resolve the H$_0$ tension. With the addition of external information from 33 Sloan Lens ACS (SLACS) lenses, the precision improves to 5\% if the deflectors of TDCOSMO and SLACS lenses are drawn from the same population. 
   We investigate the prospect of improving the precision of time-delay cosmography without relying on mass profile assumptions to break the mass-sheet degeneracy.
   Our forecasts are based on a previously published hierarchical framework.
   With existing samples and technology, 3.3\% precision on H$_0$ can be reached by adding spatially resolved kinematics of the seven TDCOSMO lenses. The precision improves to 2.5\% with the further addition of kinematics for 50 nontime-delay lenses from SLACS and the Strong Lensing Legacy Survey (SL2S).  Expanding the samples to 40 time-delay and 200 nontime-delay lenses will improve the precision to 1.5\% and 1.2\%, respectively.
   Time-delay cosmography can reach sufficient precision to resolve the Hubble tension at 3-5$\sigma$, without assumptions on the radial mass profile of lens galaxies. By obtaining this precision with and without external datasets, we will test the consistency of the samples and enable further improvements based on even larger future samples of time-delay and nontime-delay lenses (e.g., from the Rubin, Euclid, and Roman Observatories).
   }


   \maketitle
%

\section{Introduction}

Almost a century after it was first measured, the Hubble constant H$_0$ still remains arguably the most debated number in cosmology. In the past few years, a discrepancy has emerged between a number of local measurements, and inferences from early-Universe probes such as the cosmic microwave background (CMB) and Big Bang nucleosynthesis, under the assumption of flat $\Lambda$ cold dark matter ($\Lambda$CDM) cosmology \citep[see, e.g.,][for a recent summary]{Verde:2019}. If this discrepancy is real, and not due to unknown systematic uncertainties in multiple measurements, it implies that the standard $\Lambda$CDM model is not sufficient, and new physical ingredients beyond this model are required. From a theoretical standpoint, a number of possible solutions ---for example, involving early dark energy--- have been proposed \citep[e.g.,][and references therein]{KnoxMillea2020}, often requiring fine-tuning of free parameters in order to avoid violating other observational constraints. From an observational point of view, besides improving the precision of the measurements, significant attention has turned to the systematic investigation of unknown systematic uncertainties \citep[e.g.,][]{Riess:2019,Freedman:2020,Riess:2020}.

Strong lensing time delays \citep[hereafter Time-Delay Cosmography,][and references therein]{TreuMarshall2016} provide a one-step measurement of H$_0$ that is independent of any other probe, and are thus a powerful contribution to this debate. By assuming standard forms for the mass density profile of early-type galaxies ---consistent with X-ray \citep[e.g.,][]{HumphreyBuote2010} and stellar kinematic observations \citep[e.g.,][and references therein]{Cappellari2016}--- the H0LiCOW/COSMOGRAIL/STRIDES/SHARP (hereafter TDCOSMO\footnote{\url{www.tdcosmo.org}}) collaborations achieved 2\% precision on H$_0$ \citep{Rusu:2020,Wong:2020,Chen:2019,Shajib:2020,Millon:2020}, in excellent agreement with the local distance ladder measurement by the SH0ES team \citep{Riess:2019} and more than 3$\sigma$ statistical tension with early-Universe probes \citep[e.g.,][]{ACTDR4}. 
In summary, if the mass density profiles are well described by a power-law or a constant mass-to-light ratio plus a \citet{NFW} dark matter halo\footnote{Imposing standard priors on the mass and concentration of the halo.}, the tension is significant from the strong lensing measurements alone, corroborating other measurements, and new physics may be required.

Given the important implications of the above discrepancy, the TDCOSMO collaboration is performing a systematic investigation of possible systematic effects. \citet{Millon:2020,Gilman:2020} did not find any systematic uncertainty sufficient to resolve the discrepancy, if the two assumed forms of the mass density profile are valid. Attention therefore turned to relaxing the radial profile assumption. \citet[][hereafter TD4]{Birrer:2020} addressed the issue in the most direct way, by choosing a parametrization of the radial mass density profile that is maximally degenerate with H$_0$, via the mass-sheet transform \citep[MST,][]{Falco:1985}. With this more flexible parametrization, H$_0$ is only constrained by the measured time delays and stellar kinematics, increasing the uncertainty on H$_0$ from 2\% to 8\% for the TDCOSMO sample of seven lenses, without changing the mean inferred value significantly.

TD4 introduce a hierarchical framework in which external datasets can be combined with the time-delay lenses to improve the precision. These latter authors achieved a 5\% precision measurement on H$_0$ by combining the TDCOSMO lenses with stellar kinematic measurements of a sample of lenses from the Sloan Lens ACS (SLACS) survey with no time-delay information  \citep{Bolton:2008,Auger:2009}. The mean of the TDCOSMO+SLACS measurement is offset with respect to the TDCOSMO-only value, in the direction of the CMB value, although still statistically consistent given the uncertainties\footnote{The TD4 measurements are in statistical agreement with each other and with the earlier H0LiCOW/SHARP/STRIDES measurements based on radial mass profile assumptions. TD4 is also consistent, by construction, with the study by \citet{Shajib:slacs}, because they share the same measurements for SLACS. \citet{Shajib:slacs} concluded that NFW+stars (using wider priors on mass and concentration than earlier H0LiCOW/SHARP/STRIDES measurements) is a sufficiently accurate description of the mass density profile of the SLACS lenses. However, small departures from those forms are allowed by the data, resulting in the uncertainties quoted by TD4.}. The shift in the mean could be real or it could be due to an intrinsic difference between the deflectors in the TDCOSMO and SLACS samples, arising from selection effects. For example, the two samples are well matched in stellar velocity dispersion, but they differ in redshift; the TDCOSMO sample is source selected and composed mostly of quadruply imaged quasars, while SLACS is deflector selected and dominated by doubly imaged galaxies.

In this paper we outline a two-pronged strategy to improve the precision of time-delay cosmography with flexible radial mass profile assumptions, as described in TD4. We use the formalism introduced by TD4 to forecast the precision of H$_0$ attainable by improving the kinematic data of the TDCOSMO sample and by expanding and improving the kinematic data of external datasets when drawn from the same underlying deflector galaxy population. We show that, by pursuing both avenues on existing samples and with current technology, one can recover most of the precision achieved through previous stronger assumptions on the mass profile and at the same time test for internal consistency of the TDCOSMO and external datasets, thus verifying a key assumption of TD4. This dual strategy will also be beneficial in the longer term, when the sample size of both time-delay and nontime-delay lenses will expand by order of magnitudes, but the latter will always be a subset of the former due to the observational cost of measuring time delays. We stress that the point of this paper is not to discuss whether specific assumptions about the mass density profile of massive elliptical galaxies are valid or not, but rather to show that with sufficient data one can achieve 2-3\% precision on H$_0$ without making those assumptions. Of course, as a byproduct, following our proposed strategy, it is also possible to tell whether previous assumptions are sufficient to provide an accurate H$_0$ measurement.

This paper is organized as follows. In \S~\ref{sec:method} we summarize the hierarchical framework and its assumptions, referring the reader to TD4 for details. In \S~\ref{sec:future} we describe the datasets used for the forecast. In \S~\ref{sec:forecasts} we present the forecasts. In \S~\ref{sec:conc} we draw our conclusions. Our conclusions are independent of the specific value of H$_0$ chosen for the forecast. However, when necessary for visual clarity, we adopt a value of H$_0 = 70$ \Hunit and $\Omega_{\rm m}=0.3$. A standard flat $\Lambda$CDM cosmology is assumed, with a uniform prior of H$_0$ in [0, 150] \Hunit and a tight prior on $\Omega_{\rm m}$ based on relative distance measurements from type Ia supernovae with $\mathcal{N}(\mu=0.3, \sigma=0.022)$ \citep[i.e., comparable to][]{Scolnic:2018}.
The code used for the analysis presented in this work is available 
\href{https://github.com/sibirrer/TDCOSMO_forecast}{on GitHub 
}\footnote{\url{https://github.com/sibirrer/TDCOSMO_forecast}}.

\section{Summary of the hierarchical framework}
\label{sec:method}

\subsection{Background}

The MST leaves the relative imaging observables unchanged but scales the predicted time delays, posing a fundamental limitation on the power of imaging data to constrain the radial mass profile of strong gravitational lenses, and in turn, H$_0$. In terms of the convergence field, the MST describes a re-scaling of a given lens mass profile at angular coordinate $\theta$, $\kappa(\theta)$, with a factor $\lambda,$ while simultaneously adding a sheet of mass with constant convergence $(1-\lambda)$ as
\begin{equation}\label{eqn:mst}
    \kappa_{\lambda}(\theta) = \lambda \times \kappa(\theta) + \left( 1 - \lambda\right).
\end{equation}
The inferred H$_0$ value from the measured time delays scales as
\begin{equation}
    {\rm H}_{0 \lambda} = \lambda {\rm H}_0.
\end{equation}

The stellar kinematics of the deflector galaxy ---a lensing-independent mass tracer--- can constrain the MST
for a given family of mass profiles. The constraints on $\lambda$ depend on the precision of the stellar velocity dispersion ($\sigma^{\rm P}$) measurement as
\begin{equation} \label{eqn:sigma_v_lambda_error_propagation}
  \frac{\delta \lambda}{\lambda} = 2\frac{\delta \sigma^{\rm P}}{\sigma^{\rm P}}.
\end{equation}
Current uncertainties on the stellar velocity dispersion measurements of order 5\%-10\% imply that the joint analysis of time-delay and nontime-delay lens samples is required to constrain $\lambda$.
Equation \ref{eqn:sigma_v_lambda_error_propagation} does not include additional model uncertainties beyond $\lambda$ in the prediction of the velocity dispersion. The kinematic modeling generally requires a 3D de-projected mass and stellar distribution model. The measured velocity dispersion is luminosity weighted, seeing integrated, and measured in projection along the line of sight.
A key component in the interpretation of the velocity dispersion measurement, and thus the inference of $\lambda$, is the anisotropy distribution of the stellar orbits
\begin{equation} \label{eqn:anisotropy_definition}
  \beta_{\rm ani} \equiv 1 -  \frac{\sigma_{\rm t}^2}{\sigma_{\rm r}^2},
\end{equation}
where $\sigma_{\rm r}^2$ and $\sigma_{\rm t}^2$ are the radial and tangential velocity dispersions, respectively.

\subsection{Implementation of the hierarchical framework}

We adopt the framework introduced by TD4. Here we provide a brief summary for convenience referring to TD4 for details, including parametrization and adopted priors. 

The TD4 framework drastically reduces the mass profile assumptions on individual lenses with respect to previous work, and quantifies any potential effect of the MST with the MST parameter $\lambda$ applied to  a power-law radial mass density profile that is maximally degenerate with H$_0$. This approach is similar to that of \citet{Birrer:2016} who encoded the MST with a source size regularization in the inference.

Stellar velocity dispersion is assumed to be isotropic in the center and radial in the outer part, following the  \citet[][]{Osipkov:1979, Merritt:1985} form
\begin{equation} \label{eqn:r_ani}
  \beta_{\text{ani}}(r) = \frac{r^2}{\left(a_{\rm ani}r_{\rm eff}\right)^2+r^2},
\end{equation}
where $r_{\rm eff}$ is the half-light radius of the deflector and $a_{\rm ani}$ is the anisotropy scaling factor.

To account for covariances in the parameters and priors on the MST and the stellar anisotropy, TD4 introduced a hierarchical framework to describe the MST parameter $\lambda$ and the anisotropy parameter $a_{\rm ani}$ at the lens population level, assuming that the lenses are drawn from the same parent population.

The framework is validated on the Time-Delay Lens Modeling Challenge Rung3 mock lenses generated from hydrodynamical simulations \citep{Ding:tdlmc}.
An external data set of gravitational lenses with kinematics and imaging constraints can be incorporated under the assumption that the deflectors are drawn from the same population as those of the time-delay lenses. Both unresolved and spatially resolved velocity dispersion measurements can be used in this framework. The spatially resolved measurements are particularly useful to constrain the anisotropy of the stellar orbits.

\section{Future data sets}
\label{sec:future}

We envision parallel improvements in the quality of the data for the time-delay lenses in the TDCOSMO samples (\S~\ref{ssec:tdcosmo}) and external datasets composed of nontime-delay lenses (\S~\ref{ssec:external}). We consider two cases: (i) improvements that can be made with existing\footnote{We consider the James Webb Space Telescope (JWST) an existing facility, because the call for proposals for cycle-1 is open.} facilities and samples (current scenario); and (ii)\ gains that can be made with future samples and/or facilities (future scenario). The scenarios are summarized in Table~\ref{table:param_summary_tdlmc}.

A clarification is needed for the spatially resolved kinematics. There is an uncertainty floor on the  calibration of stellar velocity dispersion due to systematic effects such as the match between stellar templates and target composite stellar populations and knowledge of the instrumental properties. To account for this floor, our stated precision is the overall uncertainty on the mean of the stellar velocity dispersion across the target, while the shape of the velocity dispersion profile is constrained taking into account the covariance between spatial bins.

\subsection{Time-delay lenses}
\label{ssec:tdcosmo}

One limiting factor of the current TDCOSMO dataset is the precision of the unresolved stellar velocity dispersion measurements, which range between 5 and 10\%. A first improvement is to bring all the uncertainties to 5\%, which has been demonstrated to be feasible with ground-based spectrographs given sufficient data quality. This is the TDCOSMO-5\% scenario.

An additional improvement consists in spatially resolved stellar velocity dispersion of the TDCOSMO sample. Such data can be obtained from the ground in the optical in seeing-limited mode (e.g., with MUSE/VLT or KCWI/Keck; hereafter TDCOSMO+O-IFU), or in the infrared with adaptive optics correction (e.g., with OSIRIS/Keck; hereafter TDCOSMO+AO-IFU). JWST will enable a further improvement over ground-based spatially resolved kinematics owing to its superior stability and absence of emission and absorption from the Earth's atmosphere (hereafter TDCOSMO+JWST-IFU). In the future scenario we expand the sample to 40 time-delay lenses and assume we can use 30-m class Extremely Large Telescopes (ELTs) with adaptive optics (hereafter TDCOSMO+ELT-IFU). JWST data for this future sample would give a similar precision on H$_0$.

\subsection{External datasets}
\label{ssec:external}

There are currently three limiting factors to the external dataset used in TD4, namely (i) the precision of aperture velocity dispersion measurements; (ii) the absolute calibration and sample size of integral field data; and (iii) the overall sample size. In the current scenario, we consider two ways to overcome these limitations. The first is to take 50 lenses from the current SLACS and SL2S \citep{Sonnenfeld:2013a,Sonnenfeld:2015} samples, selected for data quality and to match the TDCOSMO sample in velocity dispersion, with unresolved velocity dispersion measured with 5\% precision (hereafter "+50"). The second involves taking 50 lenses with spatially resolved velocity dispersion measured from seeing-limited integral field data (within reach of current generation integral field spectrographs; hereafter "+50IFU"). In the future scenario, we add 200 nontime-delay lenses to the 40 time-delay lenses described above. We stress that only time delays constrain H$_0$ and therefore the external datasets have to be used in combination with the time-delay lenses.  Considering all the combinations, we are forecasting a total of 24 scenarios (see Table~\ref{table:param_summary_tdlmc}).

\begin{table*}
\caption{Observing scenarios and forecasted H$_0$ precision. We list the specifications for the different forecasts in terms of unresolved vs. resolved velocity dispersion measurements, relative precision on the velocity dispersion per lens, $\delta \sigma_{*}/\sigma_{*}$, angular resolution of the spectroscopic observation, FWHM, the radius out to where spectral information is obtained relative to the half-light radius of the deflector, R$_{\rm spec}$/R$_{\rm eff}$, and the number of radial bins in the resolved measurements, N$_{\rm bins}$. For the "current scenario", we list the percentage precision on H$_0$ for the 7 TDCOSMO-only lenses, $\delta$H$_0$, when adding 50 lenses with aperture kinematics, +50 $\delta$H$_0$, and when adding 50 lenses with IFU data, +50IFU $\delta$H$_0$. For the "future scenario", we assume 40 TDCOSMO lenses, and optionally 200 external lenses with and without IFU kinematics.}
\begin{center}
\begin{threeparttable}
\begin{tabular}{l l l l l l l l l l}
    \hline
Current scenario       & resolution & $\delta \sigma_*/\sigma_*$  & FWHM & R$_{\rm spec}$/R$_{\rm eff}$ & N$_{\rm bin}$ & $\delta H_0$ & +50 $\delta H_0$ & +50IFU $\delta H_0$ \\ 
\hline
\hline
7 TDCOSMO-5\%    & unresolved & 5\%       & $0\farcs8$ & - & 1 &8.5\%  & 7.0\%  & 2.7\%  \\ 
7 TDCOSMO+O-IFU   & resolved   & 5\%       & $0\farcs8$ & 2 & 3 &  4.7\%  & 2.9\%  & 2.6\%  \\ 
7 TDCOSMO+AO-IFU  & resolved   & 5\%       & $0\farcs1$ & 1 & 10 &4.7\%  & 3.0\%  & 2.5\%  \\ 
7 TDCOSMO+JWST-IFU  & resolved   & 3\%       & $0\farcs1$ & 2 & 10 & 3.5\%  & 2.6\%  & 2.6\%  \\ 
\hline
Future scenario       &  &  & & & & & +200 $\delta H_0$ & +200IFU $\delta H_0$ \\  
\hline
\hline
40 TDCOSMO-5\%    & unresolved & 5\%       & $0\farcs8$ & - & 1 &7.3\%  & 7.1\%  & 1.2\%  \\ 
40 TDCOSMO+O-IFU   & resolved   & 5\%       & $0\farcs8$ & 2 & 3 &  2.0\%  & 1.3\%  & 1.2\%  \\ 
40 TDCOSMO+AO-IFU  & resolved   & 5\%       & $0\farcs1$ & 1 & 10 &2.0\%  & 1.4\%  & 1.2\%  \\ 
40 TDCOSMO+ELT-IFU  & resolved   & 3\%       & $0\farcs02$ & 3 & 30 & 1.5\%  & 1.2\%  & 1.2\%  \\  

\hline
\end{tabular}
\begin{tablenotes}
\end{tablenotes}
\end{threeparttable}
\end{center}
\label{table:param_summary_tdlmc}
\end{table*}

\subsection{Limitations}

We make a few simplifications relative to TD4 to facilitate the exploration of the information content of the data sets; such simplifications are used solely to compare their statistical properties, and are not used in estimating the final uncertainties.  We assume that (i) $\lambda$ and $a_{\rm ani}$ are single-valued intrinsic distributions without scatter; (ii) line-of-sight convergence (as a contribution to $\lambda$) has zero average with a known spread of 2\%; and (iii)  all lenses have the same Einstein radius and half-light radius and thus do not incorporate an additional parametrization to encompass a potential trend in $\lambda$ as a function of projected distance from the center of the deflector. The first two assumptions do not affect our forecast precision significantly\textbf{}, considering that line-of-sight effects are only a minor contribution to the statistical error budget. A more sophisticated representation of the third effect could in principle improve the precision of the measurement based on unresolved velocity dispersion, by providing a form of spatially resolved kinematics for the ensemble, given the range in Einstein and half-light radii for the real samples\footnote{See, e.g., \citet{Padmanabhan:2004} for an example of derivation of ensemble radial profiles from integrated velocity dispersion measurements}.

Our forecasts are robust to the details of the mock samples. For completeness and repeatability we specify that we adopted uniform priors on deflector and source redshift and typical values and measurement uncertainties for the Einstein radii, effective radii, and slope of the mass density profile prior to MST, as detailed in the Jupyter Notebook\footnote{\url{https://github.com/sibirrer/TDCOSMO_forecast}}.

It is important to state some of the key simplifying assumptions of TD4, namely that (i) spherical case of the Jeans equation for kinematic modeling; and (ii) no rotational support, i.e., no bulk rotation of the lens. Most lenses are slow rotators \citep{Barnabe:2011} and therefore we expect the approximation to be valid to first order \citep[see,][for forecasts based on nonspherical kinematics]{Yildirim:2020}. 
The integral field data proposed in this paper would allow  the hierarchical framework to be expanded to include departures from spherical symmetry and pure pressure support, extend the anisotropy model, and mitigate this potential source of systematic uncertainty.

\section{Forecasts}
\label{sec:forecasts}

In examining the performance of our proposed strategies it is worth using as a reference the 2\% precision achieved under the assumption of power-law or composite radial mass profiles \citep{Wong:2020,Millon:2020}, and the 1\% precision forecasted by \citet{Shajib:2018} under the same assumption. This is the precision floor for our forecast with the current and future samples of time-delay measurements and we show that the MST can be controlled to get fairly close to this level.

\begin{figure}
  \centering
  \includegraphics[angle=0, width=\linewidth]{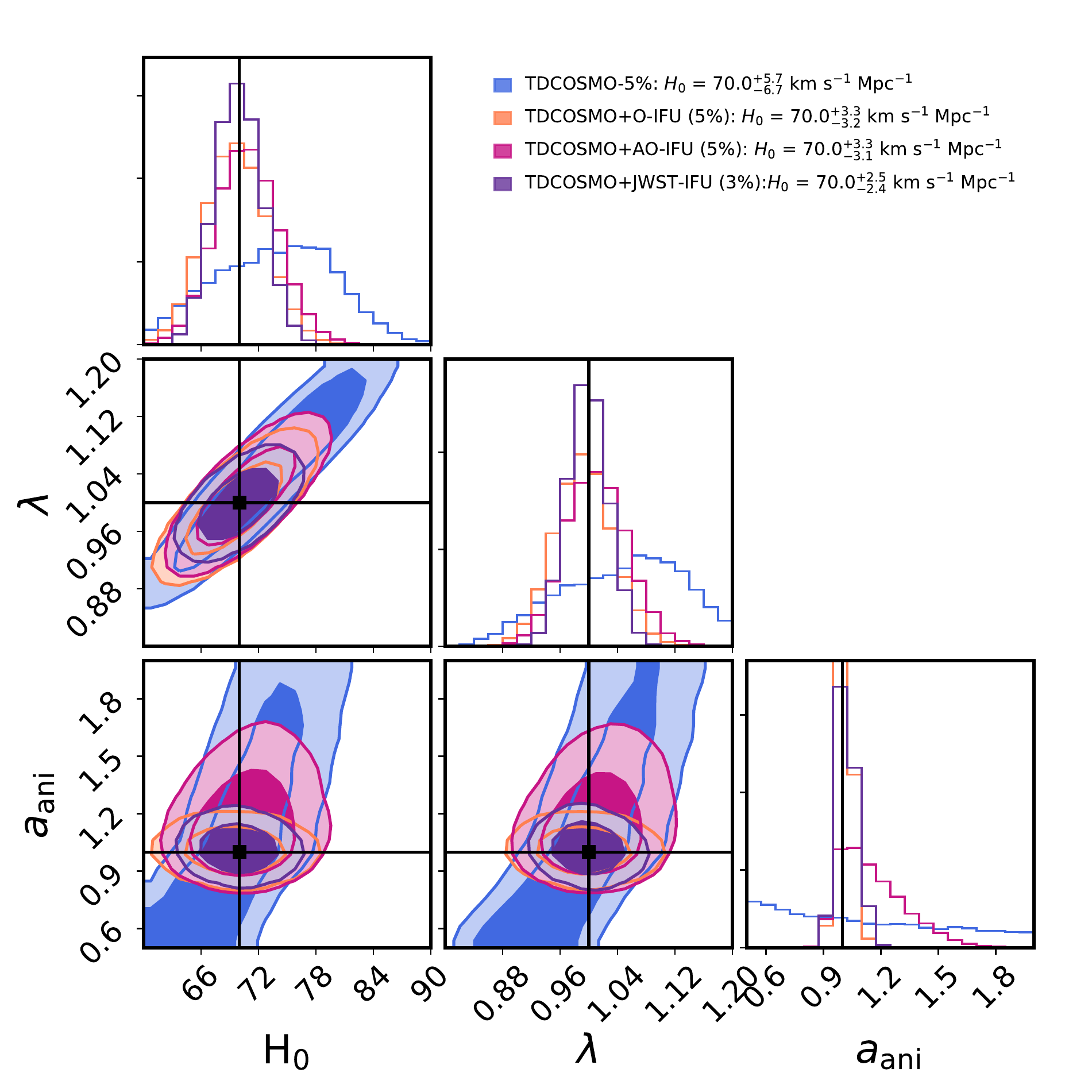}
  \caption{Forecast precision on H$_0$, the MST parameter $\lambda,$ and the anisotropy parameter $a_{\rm ani}$ for different spectroscopic scenarios of the seven TDCOSMO lenses (current scenario) as specified in Table \ref{table:param_summary_tdlmc} column $\delta$H$_0$. 
  \href{https://github.com/sibirrer/TDCOSMO_forecast/blob/master/forecast.ipynb}{~source} }
  \label{fig:tdcosmo_only}
\end{figure}

In the TDCOSMO-only current scenario (Fig.~\ref{fig:tdcosmo_only}), spatially resolved kinematics enables a precision of 3.5\% for JWST. Ground based technology reaches approximately 4.7\%, a substantial improvement over the 8.5\% without IFU, limited fundamentally by the absolute precision that can be achieved on the stellar velocity dispersion owing to instrumental effects (AO-IFU) and contamination from QSO light (O-IFU).

\begin{figure}
  \centering
  \includegraphics[angle=0, width=\linewidth]{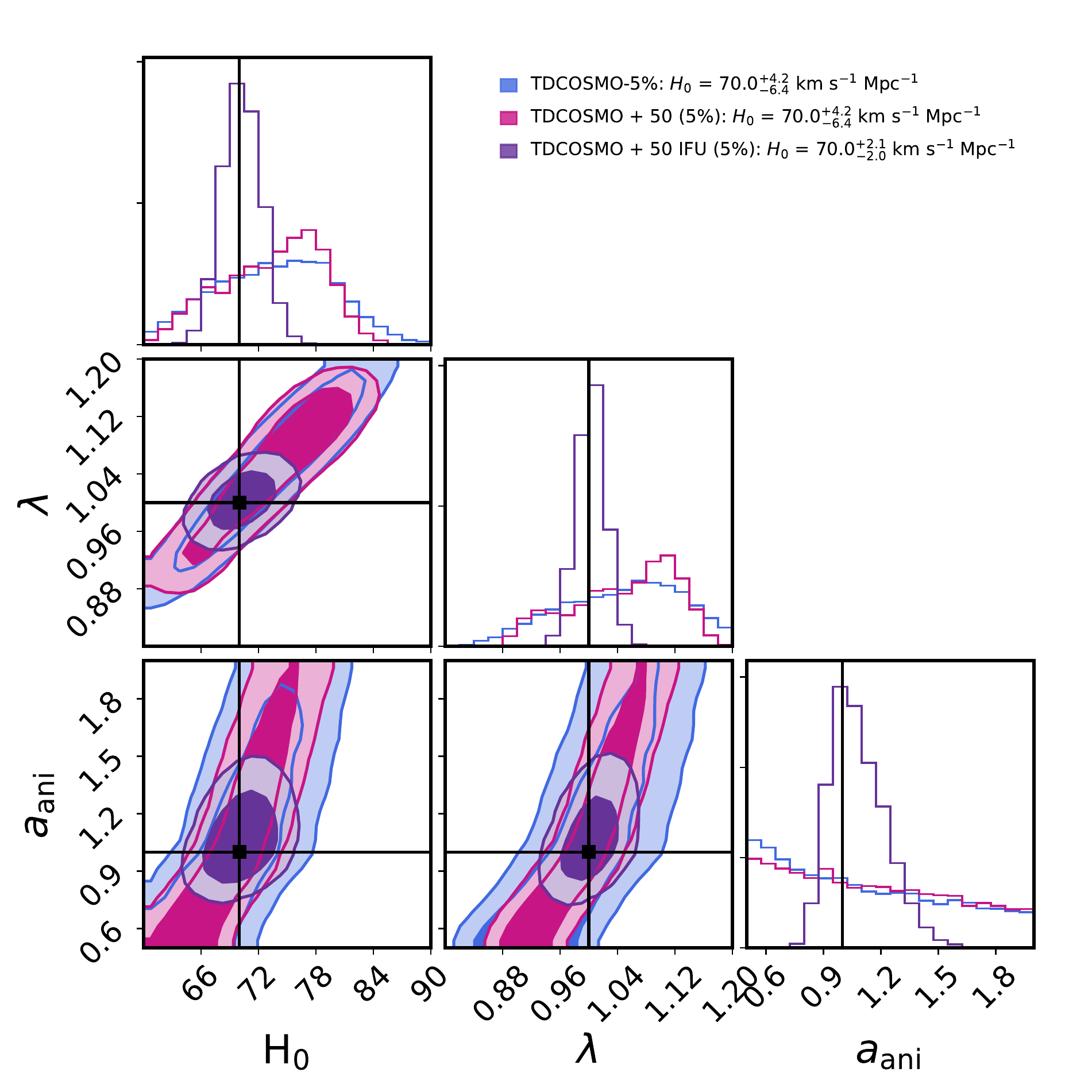}
  \caption{Forecast precision on H$_0$, the MST parameter $\lambda,$ and the anisotropy parameter $a_{\rm ani}$ for different spectroscopic scenarios of the seven TDCOSMO lenses (current scenario) observed with aperture spectroscopy of 5\% precision as well as in the case where external data sets are added, as specified in Table \ref{table:param_summary_tdlmc} (TDCOSMO-5\% row).  
  \protect\href{https://github.com/sibirrer/TDCOSMO_forecast/blob/master/forecast.ipynb}{~source} }
  \label{fig:tdcosmo_plus50}
\end{figure}

In the TDCOSMO+external current scenario (Fig.~\ref{fig:tdcosmo_plus50}), adding only unresolved velocity dispersion does not improve the precision very much because of the mass-anisotropy degeneracy \citep[e.g.,][and references therein]{Courteau2014}, and our assumption that all the lenses have the same Einstein and effective radii. However, adding IFU data breaks that degeneracy and recovers almost the same level of precision as making assumptions on the radial mass profile (2.5-2.7\% vs. 2\%). 

\begin{figure}
  \centering
  \includegraphics[angle=0, width=\linewidth]{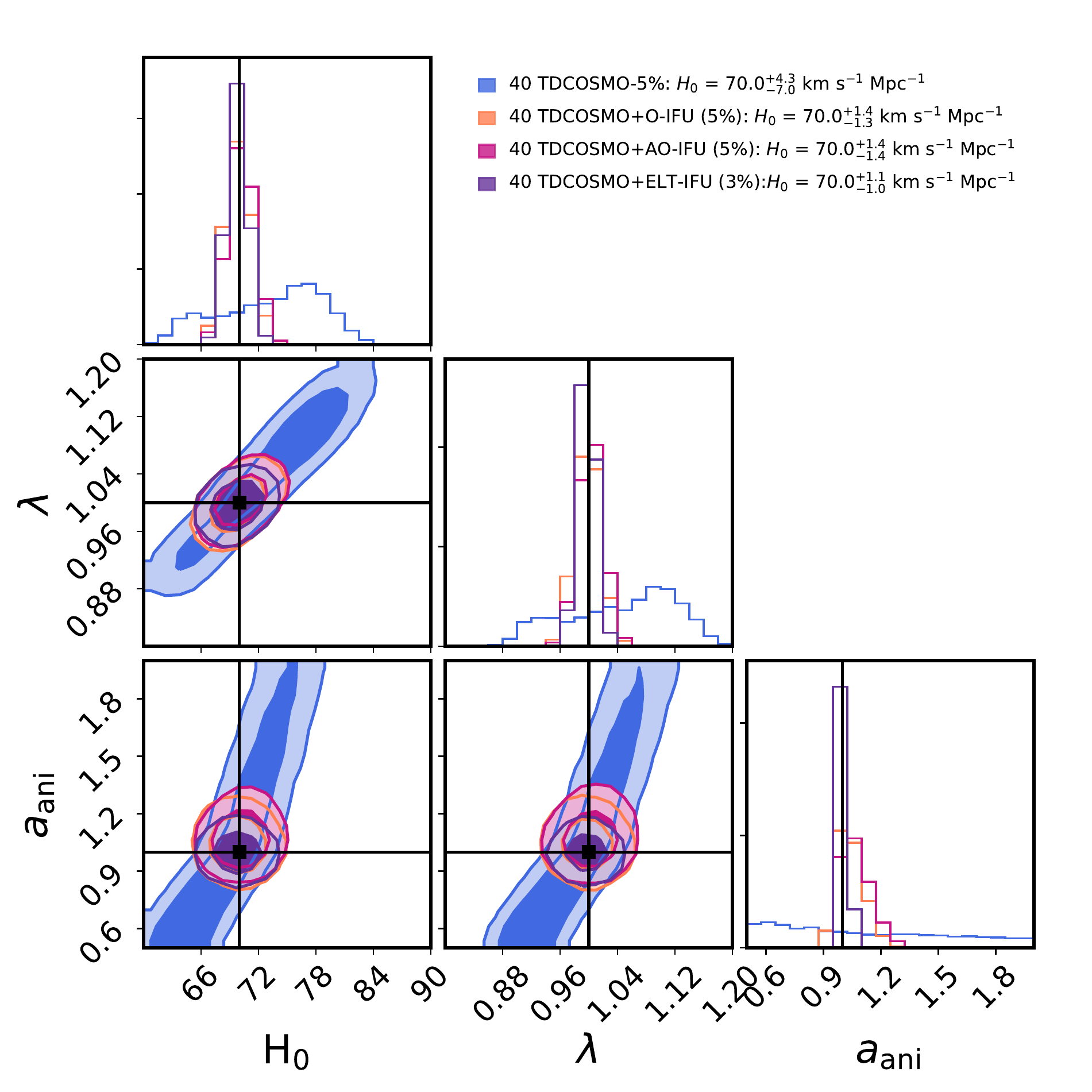}
  \caption{Forecast precision on H$_0$, the MST parameter $\lambda,$ and the anisotropy parameter $a_{\rm ani}$ for different spectroscopic scenarios of a future sample of 40 TDCOSMO lenses (future scenario) as specified in Table \ref{table:param_summary_tdlmc} in the row of TDCOSMO-5\%. 
  \protect\href{https://github.com/sibirrer/TDCOSMO_forecast/blob/master/forecast.ipynb}{~source} }
  \label{fig:tdcosmo_40_only_forecast}
\end{figure}

\begin{figure}
  \centering
  \includegraphics[angle=0, width=
 \linewidth]{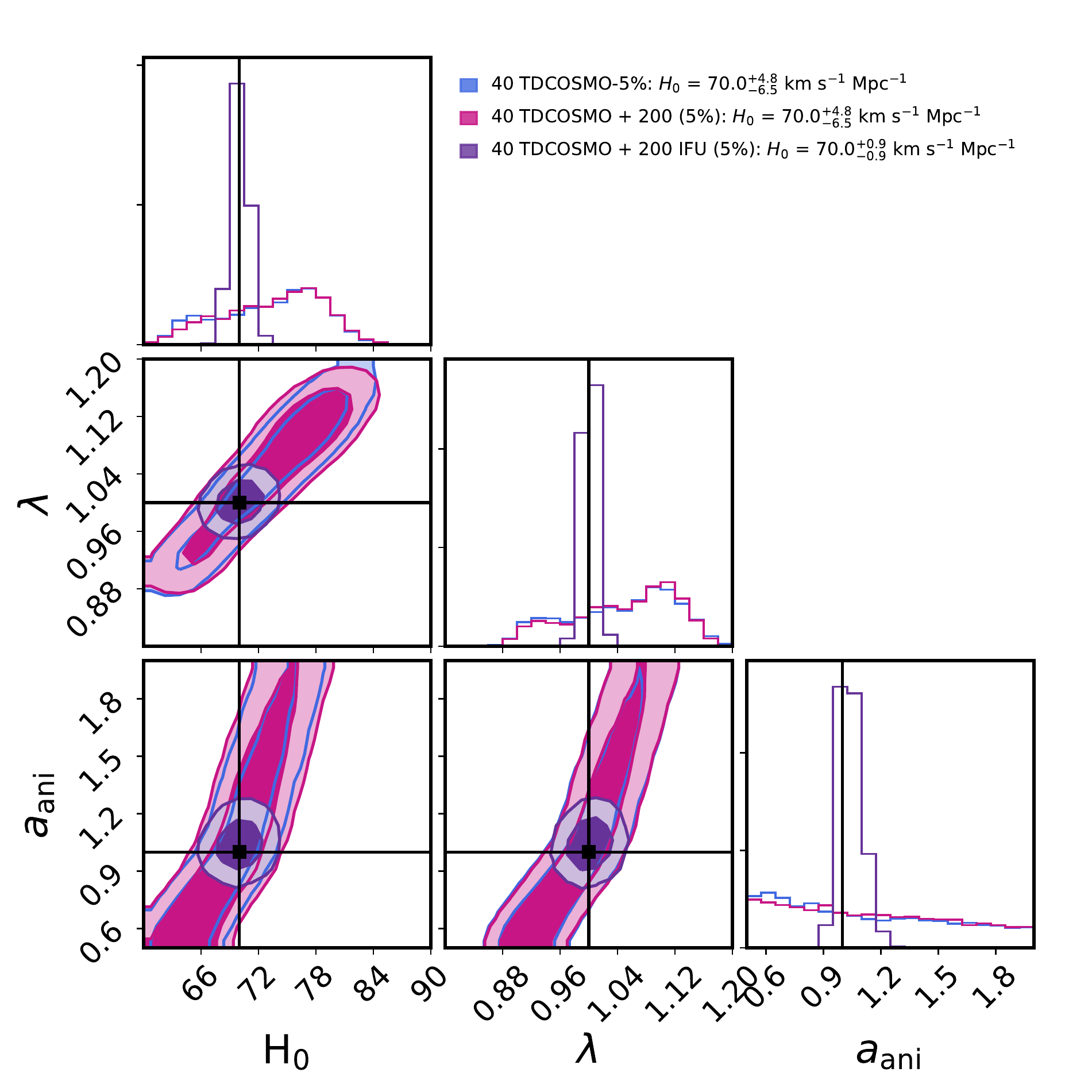}
  \caption{Forecast precision on H$_0$, the MST parameter $\lambda,$ and the anisotropy parameter $a_{\rm ani}$ for different spectroscopic scenarios of a future sample of 40 TDCOSMO lenses (future scenario) observed with aperture spectroscopy of 5\% precision and additional external data sets specified in Table \ref{table:param_summary_tdlmc} in the row of TDCOSMO-5\%. 
  \protect\href{https://github.com/sibirrer/TDCOSMO_forecast/blob/master/forecast.ipynb}{~source} }
  \label{fig:tdcosmo_plus200}
\end{figure}

We note that all the combinations that have some IFU data and at least unresolved velocity dispersion for the external dataset achieve a precision better than 3\%. (Table~\ref{table:param_summary_tdlmc}).
In this mode, the MST-related uncertainty on H$_0$ is 1.6\%, subdominant in regard to time-delay measurements, the angular lens model component, and the line-of-sight convergence of the TDCOSMO sample of seven lenses.

Similar considerations apply to the future scenarios illustrated in Figs.~\ref{fig:tdcosmo_40_only_forecast} and~\ref{fig:tdcosmo_plus200}, except for the precision that reaches 1.2-1.5\% by virtue of the larger samples. A significant component of the error budget at the 1\% level arises from the uncertainty in the relative expansion history of the Universe (in our case the prior on $\Omega_{\rm m}$).
It is encouraging that, thanks to the external datasets, we can reach a similar precision to that forecasted by \citet{Shajib:2018}. These latter authors broke the MST by assuming the mass profile is a power law. We break it with spatially resolved kinematics and external datasets.

\section{Conclusions}
\label{sec:conc}

We describe two strategies to measure H$_0$ with 2.5-3.5\% precision with gravitational time delays while accounting for the uncertainty introduced by the mass-sheet transformation. The first is based on current samples of 7 time-delay lenses and existing technology and the second is based on adding 50 nontime-delay lenses. The same strategies, applied to a future sample of 40 time-delay and 200 nontime-delay lenses can achieve 1.2-1.5\% precision. The keys to achieving this precision are spatially resolved kinematics and the inclusion of datasets of nontime-delay lenses in a hierarchical framework.

These two strategies are not mutually exclusive and both should be pursued. The TDCOSMO-only approach has the advantage of not relying on the assumption that the time-delay and nontime-delay galaxies are drawn from the same parent population. With this additional assumption, the TDCOSMO+external approach allows for further improvement in precision.  
The precision of each approach is sufficient to test the mutual consistency among different samples while simultaneously fitting for H$_0$. If verified, potentially with the extension of the hierarchical framework, the consistency will enable the cosmological exploitation of larger samples of nontime-delay lenses that are expected to be discovered by future surveys \citep{OM10}.

Following our proposed strategies, time-delay cosmography will, in the near future, have sufficient precision to distinguish the current $\sim 8$ \% difference between early and late Universe measurements at the $3-5\sigma$ level, without relying on assumptions on the radial mass profile of lens galaxies to break the mass-sheet degeneracy.

\begin{acknowledgements}
      SB and TT thank the TDCOSMO team for useful discussions, and in particular Anowar Shajib for and Frederic Courbin for internal review and final read, respectively. TT acknowledges support from the National Science Foundation through grants NSF-AST-1906976 and NSF-AST-1906976 and, from NASA through grants HST-GO-15320 and HST-GO-15652, from the Moore Foundation through grant 8548, and from the Packard Foundation through a Packard Research Fellowship.
      
      This work uses the following open-source software packages: \textsc{lenstronomy}\footnote{\url{https://github.com/sibirrer/lenstronomy}}\citep{Birrer_lenstronomy}, \textsc{hierArc}\footnote{\url{https://github.com/sibirrer/hierarc}}\citep{Birrer:2020}, \textsc{astropy} \citep{astropy:2013, astropy:2018}, \textsc{emcee} \citep{emcee}, \textsc{corner}
\end{acknowledgements}

%
\bibliographystyle{aa} 
\bibliography{BibdeskLib} 
%

\end{document}